\documentclass[final,5p,twocolumn]{elsarticle}

\journal{Journal of Nuclear Materials}

\usepackage{lineno}

\usepackage{hyperref}

\usepackage[utf8]{inputenc}

\usepackage{amsmath}
\usepackage{amsfonts}
\usepackage{amssymb}

\usepackage{color}

\newcommand{\stagei}[1] {I$_{\text{#1}}$}
\def \etal {\textit{et al.}}

\newcommand{\figscale}{0.8} 

\date{November 5, 2016}

\begin{document}

\begin{frontmatter}

\title{On the analysis of stage I in the resistivity recovery of electron irradiated iron} 

\author[a]{G. Apostolopoulos} 
\author[b,a]{Z. Kotsina}
\address[a]{Institute of Nuclear \& Radiological Sciences \& Technology, Energy \&  Safety,\\ 
N.C.S.R. ``Demokritos'', GR-153 10 Aghia Paraskevi, Greece}
\address[b] {Department of Solid State Physics, Faculty of Physics, National and Kapodistrian University of Athens, \\
Panepistimioupolis, GR-157 72 Athens, Greece}

\begin{abstract} 

The experimental results of Takaki \etal \citep{takaki1983} on the stage I resistivity recovery of electron irradiated iron are analyzed using the analytical theory of diffusion annealing formulated by Simpson \& Sossin \citep{simpson1970} and Schroeder \citep{schroeder1973} taking into account the recent first-principles calculations of Fu \etal \citep{fu2004} regarding the mobility of interstitials. 
Excellent agreement between theory and experiment is obtained by a minimal set of adjustable parameters. The results show that the diffusion annealing equations can be successfully employed for the analysis of recovery experiments in iron.

\end{abstract}

  
\end{frontmatter}

\section{Introduction}

Resistivity recovery after low temperature irradiation is one of the most sensitive experimental methods for the study of point defects in metals. In the case of iron, the detailed recovery spectra obtained by Takaki et al. \citep{takaki1983} on ultra-pure, electron irradiated specimens have contributed significantly to our understanding of the properties of self-interstitial atoms (SIA) and vacancies in this metal. 
Recently it has been possible by means of first-principles electronic structure calculations to obtain values of fundamental point defect parameters as, e.g., the migration energy of SIAs, vacancies and their clusters \cite{domain2001,fu2004}. In particular, for the SIA in iron it has been found that the most stable configuration corresponds to the $\left\langle 110 \right\rangle $ dumbbell with a migration energy of $E_m=0.34$~eV, in general agreement with experiments. Fu \etal\ \cite{fu2005} employed these parameters in combination with kinetic Monte Carlo (KMC) techniques for the simulation of the experimental recovery spectra of Takaki \etal, leading to a good agreement between theory and experiment. The comparison has greatly helped in clarifying the interpretation of the recovery spectra and the evolution of defect populations. A number of theoretical works have followed this paradigm and used the iron recovery data for the validation of their models or simulation techniques \cite{dallatorre2006,ortiz2007-a,jourdan2011}. 

Most of these theoretical studies have considered features of the recovery spectra occurring above a temperature of about $T\sim 110$~K. This is the position of recovery stage \stagei{D2} which is associated with the correlated recombination of Frenkel pairs\citep{maury1976,takaki1983,fu2005}. 
\stagei{D2} is the second part of the compound stage \stagei{D}, which is the strongest in the recovery spectrum of electron irradiated iron, comprising in total more than 50\% of the total recovery. The first part of \stagei{D}, \stagei{D1}, is centered at around $T\sim 100$~K and overlaps strongly with \stagei{D2} making their separation difficult. \stagei{D1} is generally attributed to the recombination of close Frenkel pairs. 
In the original work of Takaki \etal\ the fraction of recovery that belongs to \stagei{D1} and \stagei{D2} has been estimated only approximately by a graphical method. Nevertheless, most of the recent theoretical studies \cite{fu2005,dallatorre2006,ortiz2007-a,jourdan2011} have accepted this rough estimate as a basis for their calculations and no attempt has been made to deconvolute the two components of \stagei{D}. Such a deconvolution is needed for an accurate description of the defect evolution during the recovery experiments. 

In this paper we present a detailed analysis of the Takaki \etal\ \cite{takaki1983} stage I resistivity recovery data, based on the classical theory of diffusion annealing formulated by Simpson \& Sossin \citep{simpson1970} and Schr{\"o}der \cite{schroeder1973}. 
These authors devised a set of differential equations which describe the evolution of defect concentrations during SIA migration in stage I, taking into account both correlated and uncorrelated recombination. 
Their theory has been applied in fcc metals, achieving excellent agreement with experiments \citep{sonnenberg1972-a,thompson1975}.
Although iron, unlike other bcc metals, exhibits a stage I resistivity recovery very similar to fcc metals, an analysis based on the theory of diffusion annealing had not been attempted in previous work.
This was mainly due to the uncertainties that prevailed in earlier research regarding the behavior of SIAs in bcc metals. To interpret the available experimental results, it had been assumed that translation and rotation of the $\left\langle 110 \right\rangle $ SIA dumbbell had slightly different activation energies. The translational mode would provide only 2D planar migration while a rotation would allow the SIA to transfer to another plane, providing, thus, the 3D character of migration (see \citep{ullmaier1991} and references therein). In this context, the presence of bound close Frenkel pairs in the bcc lattice has also been anticipated, where the vacancy is situated close to, but outside the migration plane of the SIA. Recombination of such pairs is arrested until the rotational mode of the SIA is activated.
The recent \textit{ab initio} theoretical study by Fu \etal\ \citep{fu2004} suggests that in iron the activation energy of both the translational and rotational modes is effectively the same and that 3D migration of SIAs proceeds by combined translation-rotation nearest-neighbor jumps.
Thus, the theory of diffusion annealing can now be applied with increased robustness also in iron, since one of its key ingredients, the SIA migration energy, is known with great confidence from both theory and experiment \cite{fu2004,fu2005}. 
Although diffusion annealing is basically a continuous theory that does not take into account the discrete atomistic nature of point-defect reactions, its simplicity and its power to account for all relevant phenomena make it a valuable tool for the deconvolution of experimental recovery spectra.

\section{Theoretical background}
\label{sec:theory}

\subsection{Diffusion annealing}
The theory of diffusion annealing \citep{simpson1970,schroeder1973} considers the following reactions
\begin{subequations}
\begin{alignat}{2}
 \label{eq:iv}
 &\text{I} + \text{V} &&\to 0\\  
 &\text{I} + \text{I} &&\to \text{I}_2 \\
 & \dots \\
 &\text{I} + \text{I}_{k-1} &&\to \text{I}_k 
\end{alignat}
\end{subequations} 
between vacancies (V), self-interstitial atoms (I) and clusters of $k$ interstitial atoms (I$_k$). The first reaction represents the recombination of a SIA with a vacancy; the subsequent ones describe the clustering of interstitials. Only SIAs are considered to be mobile.

The corresponding defect concentrations are denoted by $n_0,\; n_1, \dots ,\; n_k$ for V, I and up to the $k$-th interstitial cluster, respectively. The temporal evolution of defect concentrations is described by a set of differential equations
\begin{subequations}
\label{eq:ss}
\begin{align}
\label{eq:ss0}
\dot{n}_0 &= -\alpha_0 n_0 n_1\\  
\dot{n}_1 &= \dot{n}_0 - n_1 ( 2 \alpha_1 n_1 + 
\sum_{k\geq 2}{\alpha_k n_k} )   \\
\dot{n}_k &= -n_1 ( \alpha_k n_k - \alpha_{k-1} n_{k-1} ), \quad k\geq 2
\end{align}
\end{subequations} 
where $\dot{n}_k$ denote derivatives with respect to time and $\alpha_k$ are the time-dependent reaction rates  
\begin{equation}
\label{eq:ak}
\alpha_k(t) = (4\pi/\Omega_0) R_k D 
\left( 1 + R_k / \sqrt{\pi (1+\delta_{1k}) D t} \right), 
\end{equation}
that were first obtained by Waite \cite{waite1957}.
In the last equation $\Omega_0$ is the atomic volume and $R_k$ denotes the reaction radius for the interaction between a SIA and a defect belonging to the concentration $n_k$. Thus, $R_0$ is the SIA-vacancy recombination radius. The SIA diffusion constant is $D=D_0 e^{-E_m/k_B T}$ with $E_m$ denoting the migration energy, $T$ the temperature and $k_B$ the Boltzmann constant. Finally, $\delta_{1k}$ denotes the Kronecker delta.

The equations \eqref{eq:ss} assume that all defect species are randomly and homogeneously distributed in space. Thus they describe the uncorrelated recombination due to long range SIA migration.

\subsection{Correlated recombination}
An essential part of diffusion annealing theory is the description of correlated recombination, i.e., the reaction of a SIA with its own vacancy. For an isolated Frenkel pair the probability of recombination after time $t$  is \citep{simpson1970}
\begin{equation}
\label{eq:pc}
p_c(t) = 
\left\langle 
\frac{R_0}{R}\text{erfc}\left( \frac{R-R_0}{\sqrt{4Dt}}\right)
\right\rangle_{g(R)},
\end{equation}
where $R$ is the initial vacancy-interstitial separation. The symbol $\langle\dots\rangle_{g(R)}$ denotes averaging\footnote{$\langle f(R) \rangle_{g(R)}=4\pi \int_{R_0}^\infty {f(R) g(R) R^2 dR} $} 
over the probability distribution $g(R)$ for the interstitial to be initially at a distance between $R$ and $R+dR$ from its vacancy. For simplicity $g(R)$ is considered as radially symmetric.
The form of $g(R)$ will largely define the experimentally observed correlated recovery. Several different expressions for $g(R)$ that lead to analytically tractable results are given in \citep{simpson1970}.

Correlated recombination is included in the diffusion equations by adding an extra term to \eqref{eq:ss0} :
\begin{equation}
\label{eq:ss02}
\dot{n}_0 = -\alpha_0 n_0 n_1 - n^0 e^{-F}
\dot{p}_c, 
\tag{\ref{eq:ss0}$'$}
\end{equation} 
where $n^0$ is the initial concentration of Frenkel pairs and $F$ is associated with the probability that an interstitial is captured by a defect other than its own vacancy. The following equation 
\begin{equation}
\label{eq:F}
\dot{F} =  \alpha_0 (n_0+n_1) + 2\alpha_1 n_1 + \sum_{k\geq 2}{\alpha_k n_k}
\end{equation}
describing the time evolution of $F$ has to be added to the eq. system \eqref{eq:ss} to complete the description.

\subsection{Integration of equations in isochronal annealing conditions}
\label{sec:eqtme}

In typical isochronal recovery experiments an irradiated sample with initial Frenkel pair concentration $n^0$ is annealed at successively increasing temperatures $T_1 < T_2 < ... < T_N$ for a specific time interval $\Delta t$ per annealing point. To obtain the evolution of defects the system of equations defined by \eqref{eq:ss}, \eqref{eq:ss02} and \eqref{eq:F} has to be integrated along the annealing intervals with initial conditions
\begin{equation}
\begin{split}
n_0 &= n_1 = n^0\\  
n_k &= 0, \quad k\geq 2.
\end{split}
\label{eq:inicond}
\end{equation} 
In each interval the reaction rates $\alpha_k$ are adjusted according to the corresponding annealing temperature. 

Since the migration energy of the SIA is considered here as known, it is convenient to multiply $t$ in the $i$-th annealing interval with the corresponding Boltzmann factor. Thus a new variable $t_i =  e^{-E_m/k_B T_i} t$ is introduced in the interval $(i-1)\,\Delta t < t \leq i\,\Delta t$. Further, to facilitate comparison with recovery experiments the concentrations $n_k$ are scaled to the initial defect concentration. 
Rewriting the equations for the scaled concentrations with $t_i$ as independent variable the rate coefficients in the $i$-th annealing interval become
\begin{equation}
\label{eq:ak2}
\begin{split}
\alpha'_k(t_i) &= \alpha_k \, n^0 \, e^{E_m/k_B T_i} = \\
& (4\pi n^0/\Omega_0) R_k D_0 
\left( 1 + R_k / \sqrt{\pi (1+\delta_{1k}) D_0 t_i} \right).
\end{split}
\end{equation}

\section{Model and fitting procedure}
\label{sec:model}

\begin{figure}[!tb]
\centering
\includegraphics[scale=\figscale]{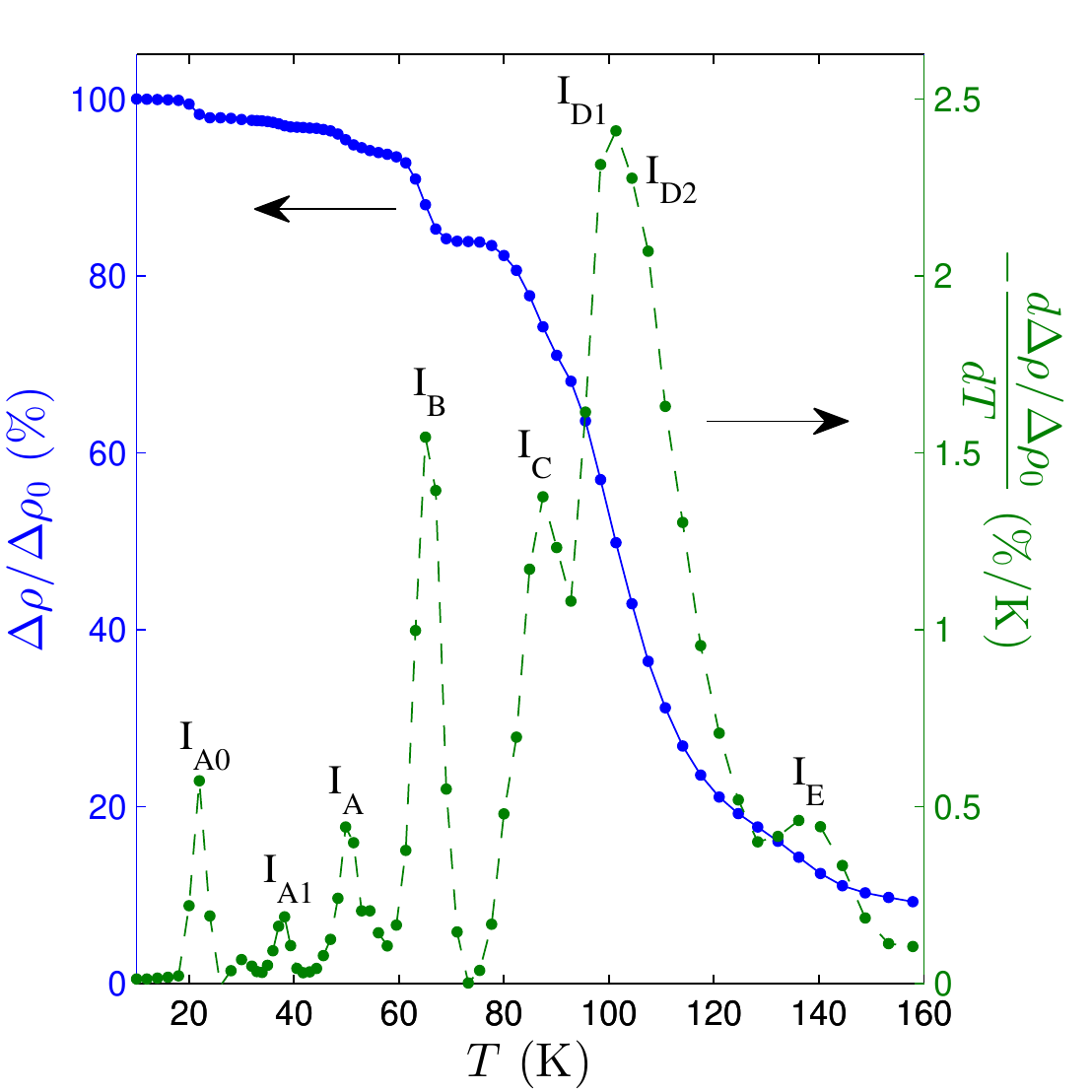}
\caption{Stage I resistivity recovery (left $y$-axis) and recovery rate (right $y$-axis) of electron irradiated Fe as a function of annealing temperature for an initial resistivity increase $\Delta\rho_0 = 22.9$~n$\Omega$-cm. Data reproduced from the work of Takaki \etal \citep{takaki1983}.} 
\label{fig:ex}
\end{figure}   

Fig. 1 shows the stage I resistivity recovery $\Delta\rho/\Delta\rho_0$ of electron irradiated iron as a function of annealing temperature for an initial resistivity increase $\Delta\rho_0 = 22.9$~n$\Omega$-cm. Also depicted on the right $y$-axis of the same figure is the resistivity recovery rate, i.e., the numerical derivative of the resistivity recovery with respect to temperature. The data has been digitally reproduced from figures 1 \& 2 of reference \citep{takaki1983}.  As seen in Fig. \ref{fig:ex} stage I consists of 8 sub-stages labeled: \stagei{A0}, \stagei{A1}, \stagei{A}, \stagei{B}, \stagei{C}, \stagei{D1}, \stagei{D2} and \stagei{E}. 

In the current work we focus on the correlated and uncorrelated Frenkel pair recombination which has been traditionally associated with \stagei{D2} and \stagei{E}, respectively. However, the lower temperature stages \stagei{C} and \stagei{D1} are also included in the analysis due to their strong overlap. 
Thus, the temperature region of interest for our analysis extends from $\sim 70$~K up to about 170~K. Takaki \etal\ \citep{takaki1983} measured the resistivity recovery at four initial dose levels, thus, the complete dataset comprises of defect resistivity values $\Delta\rho$ as a function of annealing temperature $T$ and electron dose $\phi$:
\begin{equation}
\Delta\rho(T_i,\phi_j), \quad i=1,\dots,N, \quad j=1,\dots,4.
\label{eq:dataset}
\end{equation}
The number of annealing steps $N$ is about 30. The actual electron dose $\phi$ is not explicitly specified in \citep{takaki1983}, but the initial resistivity increase $\Delta\rho_0(\phi)$ is given instead. 
Since we are not interested in the low temperature close-pair stages, we rescale the data to the value of the resistivity just before stage \stagei{C}. This corresponds to the small plateau at $T\sim 70$K, observed in the recovery curve of Fig. 1, and will be denoted by $\Delta \rho_{70}$.

In more detail the assumptions of our model are as follows:

\begin{enumerate}

\item 
At any time during stage I annealing the defect resistivity is proportional to the instantaneous concentration of Frenkel pairs $n_F$,
\begin{equation}
\label{eq:dr}
\Delta \rho = \rho_F \, n_F, 
\end{equation} 
where $\rho_F$ is the resistivity per unit pair concentration.
The typical assumptions are adopted (a) that $\rho_F$ does not depend on the distance between SIA and vacancy and (b) that the overall resistivity per defect does not change when defects take part in clusters.

\item
The total concentration of Frenkel pairs at $T=70$~K is denoted $n_{70} = \Delta\rho_{70} / \rho_F$. A fraction $f_C$ of those are close-pairs that recombine in stage \stagei{C}. The fraction $f_C$ does not depend on the initial electron dose $\phi$.

\item
The close-pair stage \stagei{C} is considered a first-order thermally activated reaction with rate constant $k_C=k_C^0 \exp(-E_C/k_B T)$, where $k_C^0$ is the pre-exponential factor and $E_C$ the activation energy. The defect evolution for such a process is well known \citep{dienes1957} and given by 
\begin{equation}
n_C(T_i,\phi_j) = n_C^0(\phi_j) \exp
\left\lbrace -\sum_{i' \leq i}{k_C(T_{i'})\,\Delta t}\right\rbrace,
\label{eq:drc}
\end{equation} 
where $n_C^0(\phi) = f_C \, n_{70}(\phi)$ is the initial concentration of close pairs.

\item 
The sub-stages \stagei{D1} and \stagei{D2} have been previously ascribed to close-pair and correlated recombination, respectively. Magnetic and mechanical relaxation measurements \citep{verdone1974,diehl1977} conducted in the temperature range of both of these stages have indicated similar activation energies of $\sim 0.3$~eV in close agreement with the results of \citep{fu2004}. Thus, in order to provide a most simple and unifying model description, we employ a single activation energy, equal to $E_m$, to both of these stages. Further, we find that the best agreement with the experimental data is achieved if both \stagei{D1} and \stagei{D2} are described by eq. \eqref{eq:pc} of correlated recombination. To account for the two peaks, the probability $g(R)$ is written as a weighted sum of two components:
\begin{equation}
g(R) = w_1 g_1(R) + (1-w_1) \, g_2(R)
\label{eq:gr2}
\end{equation} 
where $g_{1,2}(R)$ are associated with \stagei{D1,2}, respectively, and $w_1$ is a mixing parameter in the range $0\leq w_1 \leq 1$. 

The physical picture behind this assumption is the following. \stagei{D2} is indeed associated with the correlated recombination of non-close Frenkel pairs where the SIA diffuses a considerable distance before recombining with its own vacancy. On the other hand,  \stagei{D1} is due to Frenkel pairs with such an initial geometric configuration that only a very small number of SIA movements ($\sim 1 - 3$) with activation energy of $E_m$ suffice to initiate recombination. In this case we do not have a true diffusional movement of the SIA, and thus the relative distance $R$ entering $g_1(R)$ of eq. \eqref{eq:gr2} is considered an effective parameter. 

Among the possible functional forms of $g(R)$ given in \citep{simpson1970}, we find that the most suitable for the description of iron is the modified exponential   
\begin{equation}
\label{eq:gr}
g(R) = A \, e^{-R/(R_p-R_0)}/R,
\end{equation}
where $A$ is a normalization constant and the parameter $R_p$ is defined by 
$R_p^{-1} = \left\langle R^{-1}\right\rangle_{g(R)}$. 
Both $g_{1,2}(R)$ are of the type \eqref{eq:gr} but with different distribution parameters denoted $R_p^{(1,2)}$, respectively.

\item The evolution of defect concentrations in the stages \stagei{D1}, \stagei{D2} and \stagei{E} (uncorrelated recombination) are jointly described by the diffusion annealing equations of section \ref{sec:theory} using the same single SIA migration energy $E_m$. The total number of Frenkel pairs in these sub-stages is denoted as $n_D$ for simplicity, although it includes also pairs that recombine in \stagei{E}. 
By numerically integrating the diffusion equations we obtain the evolution of the pair concentration $n_D = n_{D}(T_i, \phi_j)$ as a function of annealing temperature and initial dose. The initial concentration of pairs $n_{D}^0(\phi) = (1-f_C)\, n_{70}(\phi)$ is used in place of $n^0$ in the initial conditions \eqref{eq:inicond} of the equations.

The equation system is truncated to $k\leq k_{max} = 3$, i.e., to tri-interstitial clusters. It has been found that extending to higher order clusters does not affect much the current analysis which is not concerned with the behavior above stage I. 

\end{enumerate}

According to the above statements, the following expression is used for modeling the resistivity recovery data:
\begin{equation}
\begin{split}
\frac{\Delta \rho(T_i, \phi_j)}{\Delta\rho_{70}(\phi_j)} &= 
\frac{n_C(T_i, \phi_j)+n_{D}(T_i, \phi_j)}{n_{70}(\phi_j)} \\
&= 
f_C \frac{n_C(T_i, \phi_j)}{n_C^0(\phi_j)} 
+ (1-f_C) \frac{n_{D}(T_i, \phi_j)}{n_D^0(\phi_j)}.
\end{split}
\label{eq:model}
\end{equation}

The recovery of close pairs, $n_C/n_C^0$, does not depend on the initial dose as is evident from eq. \eqref{eq:drc}. This is a well known property of first-order kinetic processes \citep{dienes1957}. Thus, the shape of \stagei{C} is the same in all dose levels and depends only on the kinetic parameters, $k_C^0$ and $E_C$. 

The situation is different for $n_D/n_D^0$, which has evidently a dose dependent behavior since the rate coefficients \eqref{eq:ak2} in the diffusion equations are proportional to the initial defect concentration. If $n^0$ is replaced in \eqref{eq:ak2} by $n_D^0=(1-f_C) \,\Delta\rho_{70} / \rho_F$  the following expression is obtained
\begin{equation}
\begin{split}
\alpha_k' &=  
 \left(\frac{4\pi R_0^3}{\Omega_0\,\rho_F}\right)
\left( \frac{R_k}{R_0} \right)
\left(\frac{D_0}{R_0^2}\right) \times \\
& \quad \left[  1 + \frac{R_k/R_0}{ \sqrt{\pi (1+\delta_{1k}) (D_0/R_0^2) t_i}} \right]\,(1-f_C)\,\Delta\rho_{70},
\end{split}
\label{eq:ak3}
\end{equation}  
where the various parameters have been rearranged using $R_0$ as a length scale. 
A similar rearrangement can also be done in the expression of correlated recovery, eq. \eqref{eq:pc}, thus the recombination radius $R_0$ is used globally as a length scale. 
The ratio $R_0^3/\rho_F$ in the right hand side of \eqref{eq:ak3} is considered a fixed constant and its value is adopted from previous experimental results \citep{dural1977}. Table \ref{tbl:fixpar} summarizes all parameters that are fixed during the analysis. 
These include also the higher order interaction radii $R_k$, $k=1,2$, which are set as slightly larger than $R_0$ in accordance with recent simulations \citep{ortiz2007-a}. However, the exact values of $R_k/R_0$ are actually not so important for the current analysis. Summarizing the parametrization of the model for the sub-stages \stagei{D} and \stagei{E}, the adjustable parameters are: the scaled pre-exponential factor $D_0/R_0^2$ and the three parameters that define $g(R)$, namely, $R_p^{(1)}/R_0$, $R_p^{(2)}/R_0$ and $w_1$.
\begin{table}[tb!]
\centering
\begin{tabular}{lccc}
\hline\hline
$E_m$	  & (eV) 		  &	0.34 	& \citep{fu2004}\\
$\frac{4}{3}\pi\, R_0^3/(\Omega_0\,\rho_F)$  & ($\Omega$-cm)$^{-1}$ &	$1.0\times 10^5$  & \citep{dural1977}\\
$R_1/R_0$ & {} 			  &	1.07 	& \citep{ortiz2007-a}\\
$R_2/R_0$ & {} 			  &	1.11	& \citep{ortiz2007-a} \\
\hline\hline
\end{tabular} 
\caption{Values of fixed parameters}
\label{tbl:fixpar}
\end{table}

The adjustable parameters are obtained by minimization of the sum-of-squares
\begin{equation}
\sum_{i=1}^N{\sum_{j=1}^4{\left\lvert 
\left[ \frac{d}{dT}\frac{\Delta\rho(T_i,\phi_j)}{\Delta\rho_{70}(\phi_j)}\right]_{\text{exp}} - 
\left[ \frac{d}{dT}\frac{\Delta\rho(T_i,\phi_j)}{\Delta\rho_{70}(\phi_j)}\right]_{\text{th}}
\right\rvert^2}}.
\label{eq:chi2}
\end{equation}
The model is fitted to the recovery rate data to achieve highest sensitivity on the position and magnitude of individual sub-stages. All four dose levels are fitted simultaneously. This makes possible the reduction of the overall number of adjustable parameters and improves the robustness and statistical variance of the results. 

The numerical integration of the diffusion annealing equations and the least-square minimization procedure have been implemented in the OCTAVE computing environment \citep{eaton2015}. The relevant programs are available at the website of the authors' institution \citep{apostolopoulos2015}.

\section{Results}

\begin{figure}[!tb]
\centering
\includegraphics[scale=\figscale]{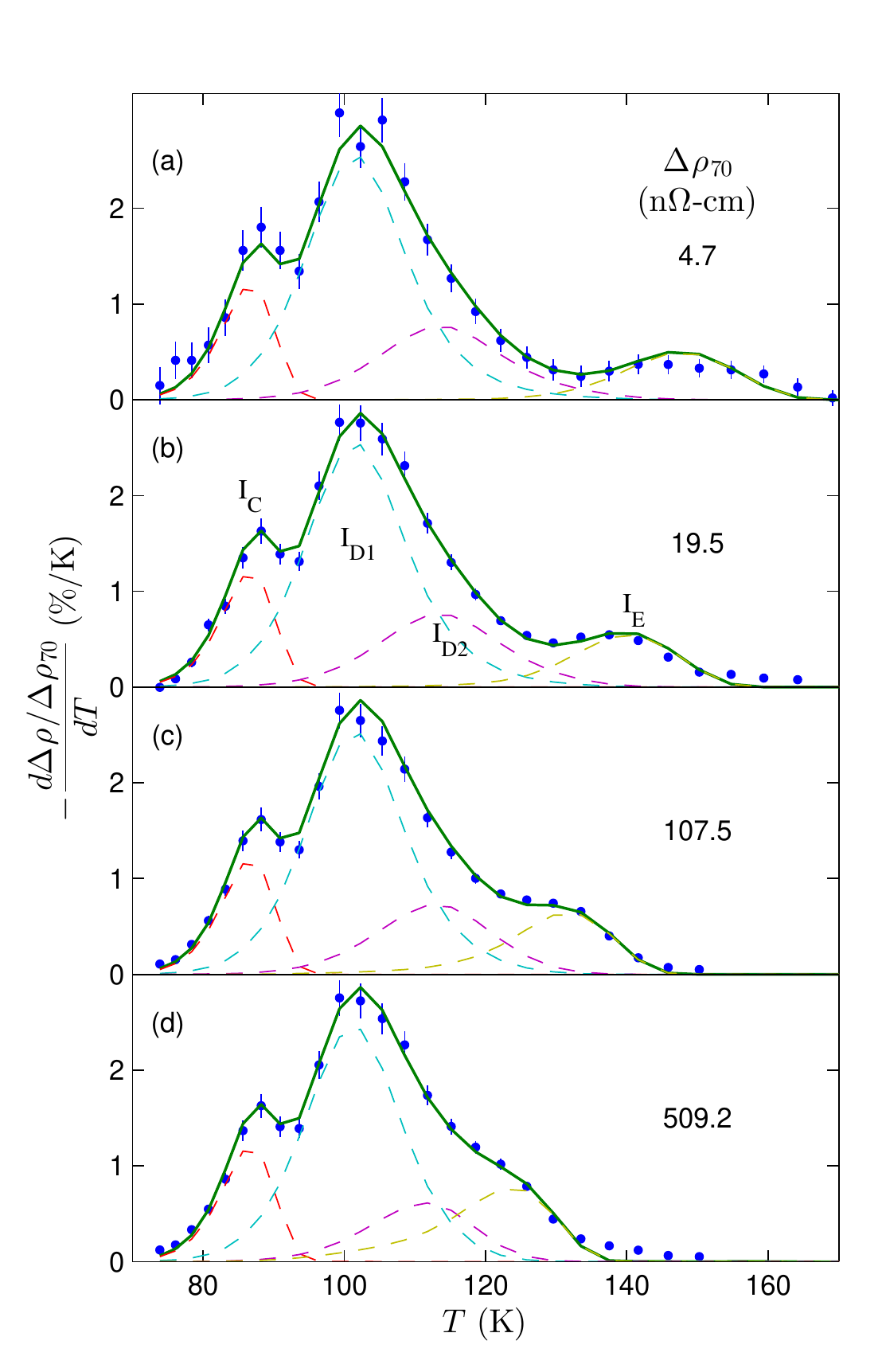}
\caption{Results of fitting the model described in section \ref{sec:model} to the  resistivity recovery rate of electron irradiated Fe. The data of Takaki \etal \citep{takaki1983} as a function of annealing temperature are denoted by dots. Sub-figures (a)-(d) refer to increasing initial electron dose. 
The fitted model is shown as a continuous curve. The dashed curves depict the model contributions of the individual sub-stages \stagei{C}, \stagei{D1}, \stagei{D2} and \stagei{E} as labeled in sub-figure (b).} 
\label{fig:fit}
\end{figure} 

Fig. \ref{fig:fit} shows the results of fitting the diffusion annealing model described in the preceding sections to the resistivity recovery rate measured in electron irradiated iron in the temperature range $70 \, \text{K} \leq T \leq 170 \, \text{K}$. The experimental recovery rate, scaled to the defect resistivity at $70$~K, is depicted as dots. The four electron dose levels measured by Takaki \etal\ are given in the sub-figures (a)-(d) and the corresponding values of $\Delta\rho_{70}$ are denoted on the graphs. Error bars are estimated by the experimental uncertainty reported in \citep{takaki1983} regarding the measurement of resistivity and by assuming a temperature error of $\pm 0.1$~K. The recovery rate predicted by the model is shown by the continuous curves. The dashed curves indicate the contributions from the different sub-stages \stagei{C}, \stagei{D1}, \stagei{D2} and \stagei{E} as labeled for example in fig. \ref{fig:fit}(b). As observed in the figure, there is good agreement between theoretical model and experimental results. It is noted that all curves are produced by a single set of parameters that capture the behavior in the whole range of initial defect concentrations.
The fitted model parameters are given in Table \ref{tbl:results}. Parameter errors refer to the statistical uncertainty of the least-square estimation procedure.

\begin{table}[!tbp]
\centering
\begin{tabular}{ccc}
\hline\hline
$f_C$ & (\%) & 11.6$\pm$0.8\\
$k_C^0$ & (s$^{-1}$) & $10^{8 \pm 1}$ \\
$E_C$ 	& (meV)	& $185\pm 10$ \\
\hline
$D_0/R_0^2$ & (s$^{-1}$) & $(1.0\pm0.2)\times 10^{12}$ \\
$R_p^{(1)}/R_0$ & {} & $1.11\pm 0.01$\\
$R_p^{(2)}/R_0$ & {} & $1.90\pm 0.15$\\
$w_1$ & (\%) & $61\pm 4$ \\
\hline\hline 
\end{tabular} 
\caption{Results of fitted model parameters.}
\label{tbl:results}
\end{table}

The first three parameters of Table \ref{tbl:results} refer to the close-pair stage \stagei{C}. The shape of the sub-stage is well described by a first-order kinetic law as seen in fig. \ref{fig:fit}. The fitted kinetic parameters $k_C^0$ and $E_C$ listed in Table \ref{tbl:results} are in excellent agreement with those reported in a previous study \citep{wells1976}.
The relative contribution of \stagei{C} to the recovery is set by $f_C$ which is found equal to 11.6 \%. 

\begin{figure}[!tb]
\centering
\includegraphics[scale=\figscale]{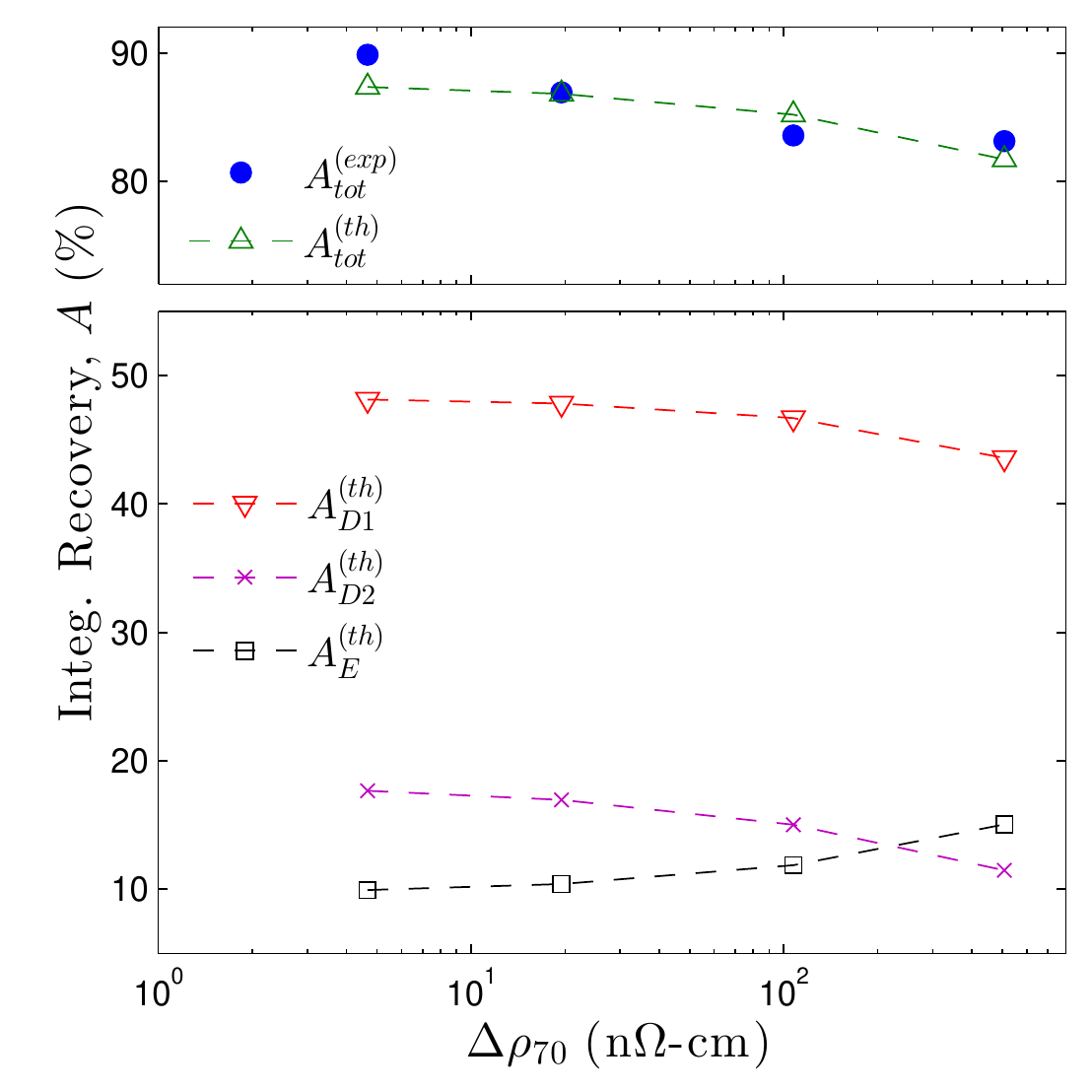}
\caption{Integrated resistivity recovery $A$ in the studied temperature range 70 - 170~K as a function of the initial resistivity increase at 70~K, $\Delta\rho_{70}$. The experimentally observed total recovery is depicted with the full dots. The results of the theoretical model for the total recovery ($\vartriangle$) and the recovery in the sub-stages \stagei{D1} ($\triangledown$), \stagei{D2} ($\times$) and \stagei{E} ($\square$) are also shown. Dashed curves are guides to the eye. }
\label{fig:A}
\end{figure}

The model accounts correctly for the observed total amount of recovery in the studied temperature range. To demonstrate this in more detail we plot in Fig. \ref{fig:A}, as a function of $\Delta\rho_{70}$, the total integrated recovery $A^{(exp)}_{tot}$, as observed experimentally, in comparison with the integrated recovery $A^{(exp)}_{th}$, obtained by the theoretical model. $A^{(exp)}_{tot}$ is calculated by
\begin{equation}
A^{(exp)}_{tot}(\phi_j) = 1- 
\left[ 
\Delta\rho(T_{max},\phi_j)/\Delta\rho_{70K}(\phi_j)
\right]_{exp},
\end{equation}
where $T_{max}$ is the maximum temperature studied at each dose level ranging between 150 and 170~K. As observed in Fig. \ref{fig:A}, 
$A^{(exp)}_{tot}$ starts at $\sim 90 \%$ at low dose and then gradually reduces to about $83 \%$ at high dose. The magnitude of the total recovery and its behavior as a function of initial dose is correctly reproduced by the diffusion annealing theory, which is also shown in Fig. \ref{fig:A}. 
Looking at the integrated recovery due to the individual sub-stages, it is observed that the reduction in total recovery with increasing dose is mainly associated with the decrease of $A_{D1}$ and $A_{D2}$. This decrease is partially compensated by a simultaneous increase of $A_E$. 
This behavior can be understood as follows. 
At higher initial Frenkel pair concentrations there is an increased probability for a SIA to escape correlated recombination, either by recombining with a vacancy other than its own or by reacting with another SIA or SIA cluster. In both cases $A_D$ is reduced. However, in the first case, when the SIA recombines with a vacancy, $A_E$ is correspondingly increased so that the total recovery remains effectively unchanged. Thus the observed reduction of total integrated recovery with dose is solely due to the enhancement of SIA clustering.

Regarding the sub-stages \stagei{D1} and \stagei{D2}, the data of Table \ref{tbl:results} show that a consistent description of the experimental data is offered if the major fraction $w_1=61\%$ of Frenkel pairs is characterized by a parameter $R_p^{(1)}=1.11R_0$ and the rest by a relatively larger value $R_p^{(2)}=1.9R_0$. 
The corresponding features in the calculated recovery spectra of fig. \ref{fig:fit} are centered at $T=103$ and 115~K, respectively. 
The value of $R_p^{(1)}$  indicates that a large number of SIAs are initially situated in the immediate vicinity of the recombination radius. For a better understanding of the underlying recovery process it is useful to estimate the average number of jumps of such a SIA before it recombines with its own vacancy. 
To obtain this average number of jumps we consider the following simple argument.
According to eq. \eqref{eq:pc}, which is employed in the description of \stagei{D1} and \stagei{D2}, the fraction of Frenkel pairs that finally recombine with their own vacancy at $t \to \infty$ is $p_c(\infty) = \langle R_0/R \rangle = R_0/R_p$ \citep{simpson1970}. Half of these pairs have recombined after time $t_{1/2}$ such that $p_c(t_{1/2})=(1/2) p_c(\infty)$, or, from eq. \eqref{eq:pc}, 
\begin{equation}
\left\langle 
\frac{R_0}{R}\text{erfc}\left( \frac{R-R_0}{\sqrt{4Dt_{1/2}}}\right)
\right\rangle_{g(R)} = \frac{1}{2} \frac{R_0}{R_p}.
\label{eq:cond}
\end{equation}  
To obtain $t_{1/2}$ from \eqref{eq:cond} to a first approximation, $R$ inside the brackets $\langle\dots\rangle_{g(R)}$ is replaced by $R_p$. Further, noting that the solution of $\text{erfc}(x)=1/2$ is approximately $x\approx 1/2$, eq. \eqref{eq:cond} simplifies to $Dt_{1/2} \approx (R_p - R_0)^2$. 
The number of nearest neighbor jumps of a SIA dumbbell migrating in the bcc lattice is $N=8 D t/a^2$, where $a$ denotes the lattice constant (\citep{was2007}, ch. 4). Thus, the number of SIA jumps corresponding to $t_{1/2}$ is $N_{1/2} = 8 (R_0/a)^2 (R_p/R_0 - 1)^2$. 
$N_{1/2}$ has been evaluated using the $R_p$ values from Table \ref{tbl:results} and the value of $R_0/a=3.3$ employed in recent calculations \citep{fu2005, ortiz2007-a}. The results are given in Table \ref{tbl:nhalf}. 
As seen from the Table, the SIA defects corresponding to \stagei{D1} recombine with their vacancy after only one jump, on the average. 
This justifies our initial assumptions regarding \stagei{D1}, namely that it corresponds to a close pair configuration that recombines after just a small number of SIA jumps. We note, however, that such a process is only approximately described by eq. \eqref{eq:pc} which has been derived on the basis of the macroscopic diffusion equation \cite{simpson1970}. The results of the present analysis must be viewed taking this limitation into account.
In \stagei{D2} the average number of jumps is about 70, thus, in this case, the SIA performs a true diffusive motion before recombination with its vacancy.   

\begin{table}[!tbp]
\centering
\begin{tabular}{ccc}
\hline\hline
{} & \stagei{D1} & \stagei{D2} \\
$N_{1/2}$  & $1.0\pm 0.2$ &  $70 \pm 20$\\
\hline\hline 
\end{tabular} 
\caption{Average number of jumps $N_{1/2}$ for SIA recombination with its own vacancy in sub-stages \stagei{D1} and \stagei{D2}.}
\label{tbl:nhalf}
\end{table}

Our results are in broad agreement with the work of Terentyev \etal\ \citep{terentyev2012} who used KMC and molecular dynamics methods to simulate the resistivity recovery experiments reported in \citep{abe1999}. These authors also attributed \stagei{D1} and \stagei{D2} to the distribution of Frenkel pairs with respect to $R$ and they also found that a large fraction of pairs have initial separation distance very close to $R_0$. However, the irradiation experiments they simulated were performed at 77~K and thus the low temperature part of \stagei{D1} and its overlap with \stagei{C} have not been studied in detail. 

It is noted that the parameters $R_p^{(1,2)}$ are not only related to the shape and temperature position of the recovery features but define also the integrated recovery in \stagei{D1} and \stagei{D2}, which is equal to $\langle R_0/R \rangle_{g(R)}$ \citep{simpson1970}. The value of this quantity as obtained here is $\langle R_0/R \rangle= R_0 [ w_1/R_p^{(1)}+(1-w_1)/R_p^{(2)} ]=0.76\pm 0.02$. 
This is lower than the value reported by Takaki \etal\ \citep{takaki1983}, who gave an estimate of $\langle R_0/R \rangle \approx 3.3/4 = 0.825$. It is also noted that $\langle R_0/R \rangle$ is related to the integrated recovery shown in Fig. \ref{fig:A} by the relation $\left( A_{D1} + A_{D2} \right) _{\phi\to 0}=(1-f_C)\,\langle R_0/R \rangle=67\%$.

Regarding finally the uncorrelated stage \stagei{E}, there is excellent agreement between our model calculation and the experimental results as observed in Fig. \ref{fig:fit}, not only with respect to the temperature position of the maximum as a function of dose but also in the relative magnitude and shape of the stage. The model reproduces very well the gradual merging of \stagei{E} into the high temperature side of \stagei{D} with increasing dose and the associated change of peak shape due to the competition between correlated and uncorrelated recombination. The key adjustable parameter associated with \stagei{E} is $D_0/R_0^2$, which is found equal to $10^{12}$~s$^{-1}$. We note that the same parameter is also involved in correlated recombination, since it actually sets the speed of SIA diffusion. Thus, its value is adjusted so that a consistent description is achieved for both \stagei{D} and \stagei{E}. The result obtained here agrees well with values employed in recent simulations \citep{fu2005, ortiz2007-a}, namely $D_0/a^2=10^{13}$~s$^{-1}$ and $R_0/a=3.3$ that correspond to a $D_0/R_0^2$ of $0.9 \times 10^{12}$~s$^{-1}$. 

Finally, we note that due to the parametrization adopted in the present model the obtained results are independent of the exact values of $R_0$ and $\rho_F$, which are both still not known with good accuracy from theory or experiment \citep{broeders2004,nakashima2015}.

\section{Conclusions}

We employ the theory of diffusion annealing formulated by Simpson \& Sossin \citep{simpson1970} and Schroeder \citep{schroeder1973} to analyze the experimental results of Takaki \etal \citep{takaki1983} on the stage I resistivity recovery of electron irradiated iron. The theory gives an accurate account of the evolution of Frenkel defect concentration as a function of temperature and dose under the following main assumptions
\begin{itemize}

\item The recovery sub-stages \stagei{D1}, \stagei{D2} and \stagei{E} are characterized by a single activation energy, $E_m$. This activation energy is taken to be equal to 0.34 eV according to recent first principles calculations \citep{fu2004}.

\item 
\stagei{D1} and \stagei{D2} are both described by the equations of correlated recombination. However, in \stagei{D1} the pairs have very short initial interstitial-vacancy separation distance $R$ and recombine after very few SIA jumps. 
This fact and the value of activation energy support the notion of \stagei{D1} as a substage where recombination of bound close pairs takes place.
In \stagei{D2} the distance $R$ of Frenkel pairs is larger and a true diffusive motion of the SIA is required for recombination. 
\end{itemize}

From the analysis we obtain refined values for a number of parameters including the pre-exponential factor for the Arrhenius law of SIA diffusion and the parameters characterizing the distribution Frenkel defects that are in good agreement with recent theoretical results.

The successful application of the theory of diffusion annealing to the results of Takaki \etal\ shows that it constitutes a suitable and powerful tool for the analysis of recovery experiments in iron.

\section*{Acknowledgments}

The authors would like to thank S. Messoloras for fruitful discussions. This work has been supported by the European Communities under the contract of Association between EURATOM and the Hellenic Republic and was carried out within the framework of the European Fusion Development Agreement (EFDA). The views and opinions expressed herein do not necessarily reflect those of the European Commission.

\bibliographystyle{elsarticle-num}
\bibliography{Paper}

%
%

\end{document}